\documentclass[aps,prd,twocolumn,eqsecnum,showpacs,preprintnumbers,superscriptaddress]{revtex4}
\usepackage{graphicx}

\newcommand{\be}{\begin{equation}}
\newcommand{\ee}{\end{equation}}
\newcommand{\ba}{\begin{eqnarray}}
\newcommand{\ea}{\end{eqnarray}}

\newcommand{\nn}{\nonumber}

\newcommand{\Li}{\mathop{\mathrm{Li}}\nolimits}
\newcommand{\Si}{\mathop{\mathrm{S}}\nolimits}
\newcommand{\re}{\mathop{\mathrm{Re}}\nolimits}

\newcommand{\vp}{\varphi}

\begin{document}

\preprint{DESY~09--070\hspace{13.4cm} ISSN 0418--9833}

\title{Heavy-quark pair production in polarized
photon-photon collisions at next-to-leading order:
Fully integrated total cross sections}


\author{B.A.~Kniehl}
\email{kniehl@desy.de}
\affiliation{II. Institut f\"{u}r Theoretische Physik,
Universit\"{a}t Hamburg, Luruper Chaussee 149, 22761 Hamburg, Germany}

\author{A.V.~Kotikov}
\thanks{On leave from Bogoliubov Laboratory for Theoretical Physics, JINR,
141980 Dubna (Moscow region), Russia.}
\email{kotikov@mail.desy.de}
\affiliation{II. Institut f\"{u}r Theoretische Physik,
Universit\"{a}t Hamburg, Luruper Chaussee 149, 22761 Hamburg, Germany}

\author{Z.V.~Merebashvili}
\email{zakaria.merebashvili@desy.de}
\affiliation{II. Institut f\"{u}r Theoretische Physik,
Universit\"{a}t Hamburg, Luruper Chaussee 149, 22761 Hamburg, Germany}

\author{O.L.~Veretin}
\email{veretin@mail.desy.de}
\affiliation{II. Institut f\"{u}r Theoretische Physik,
Universit\"{a}t Hamburg, Luruper Chaussee 149, 22761 Hamburg, Germany}

\date{\today}

\begin{abstract}
We consider the production of heavy-quark pairs in the collisions of
polarized and unpolarized on-shell photons and present, in analytic form, the
fully integrated total cross sections for total photon spins $J_z=0, \pm 2$ 
at next-to-leading-order in QCD.
Phenomenological applications include $b\bar b$ production, which 
represents an irreducible background to standard-model
intermediate-mass Higgs-boson production, as well as $t\bar t$ production.
\end{abstract}


\maketitle

\section{\label{intro}Introduction}

It has been emphasized by many physicists that running a future $e^+e^-$
linear collider
(ILC) in the
photon-photon mode is a very interesting option (see, e.g.,
Refs.~\cite{proc95, badelek04}). The high-energy on-shell
photons can be generated by backward Compton scattering of laser light off the
high-energy electron and positron bunches of the collider with practically no
loss in energy and luminosity.
In this respect, one of the most important reactions to consider is
heavy-quark pair
production in photon-photon collisions. A $\gamma\gamma$ collider becomes
particularly important for studies of the standard-model Higgs boson when
its mass is below the $W^+ W^-$ production threshold.
Then, the predominant decay is $H\to b\bar b$.
The dominant background to this comes from $\gamma\gamma \to b\bar b$, which
receives contributions from direct and resolved photons.
We leave aside the latter for the time being and return to this in
Sec.~\ref{summary}.
The use of longitudinally
polarized photons of equal helicity (their angular momentum being $J_z=0$)
suppresses this background by
a factor of $m_b^2/s$ at the leading order in perturbation theory
\cite{Haber, Borden}.
Of course, the reason that the $J_z=0$ channel is important is that the Higgs
signal comes entirely from it.
Nevertheless, the above-mentioned suppression should not
necessarily hold in general, since QCD higher-order corrections involve
gluon emission, which permits the $b\bar b$ system to have $J_z \ne 0$.
Therefore, the process of bottom-quark pair production in
polarized-photon fusion would represent an irreducible background to
intermediate-mass Higgs-boson production.
Indeed, subsequent calculations of the next-to-leading-order (NLO) QCD
corrections have confirmed these expectations \cite{KMC, Jikia}.

Furthermore, future photon colliders will become top-quark factories.
The data obtained there, when combined with data on top-quark production 
from other reactions,
will certainly improve our knowledge of the top-quark properties (see, e.g.,
Ref.~\cite{hewett98}). It should also be noted that the NLO corrections
have a large effect on the threshold behavior and exhibit a peculiar 
spin dependence in this region.

Heavy-flavor production in photon-photon collisions receives 
contributions from direct and resolved incident photons.
In the first case, photons behave as pointlike objects, interacting
directly with the quarks in the hard scattering, 
while in the second case, the photon exhibits a complex structure involving 
quarks and gluons that participate in the hard interaction.
In this paper, we present analytical results for the total cross sections
for heavy-quark pair production by both polarized and unpolarized direct
photons.
The present work builds on the previous work of one of us \cite{KMC, KM}.
In Ref.~\cite{KMC}, differential cross sections were calculated
analytically in dimensional regularization \cite{DREG} and cast into a very
compact form.
We note that this is the only publication where complete analytical
results for polarized and unpolarized doubly differential 
cross sections are presented.
In Ref.~\cite{KM}, top-quark pair production for energies not too far above
threshold was studied, and the fully integrated result for the
so-called ``virtual plus soft'' part of the cross
section was derived.
We also note that the results presented in the present work constitute
the Abelian part of the gluon-induced hadroproduction of
heavy-quark pairs.

This paper is organized as follows. Section~\ref{notation} explains
our notations.
In Sec.~\ref{evaluation}, we 
outline our general approach and 
discuss in detail our procedure and methodology.
In Sec.~\ref{results}, we present our analytically integrated
total cross sections.
Our conclusions are summarized in Sec.~\ref{summary}. Finally,
Appendix~\ref{I8integral} elaborates on the calculation of one of the most
difficult double integrals, 
Appendix~\ref{coeffs} gives expressions for the various coefficient
functions that appear in the main text, 
and Appendix~\ref{functions} displays 
representations of our basis functions in terms of generalized
Nielsen polylogarithms.

\begin{figure*}
\includegraphics{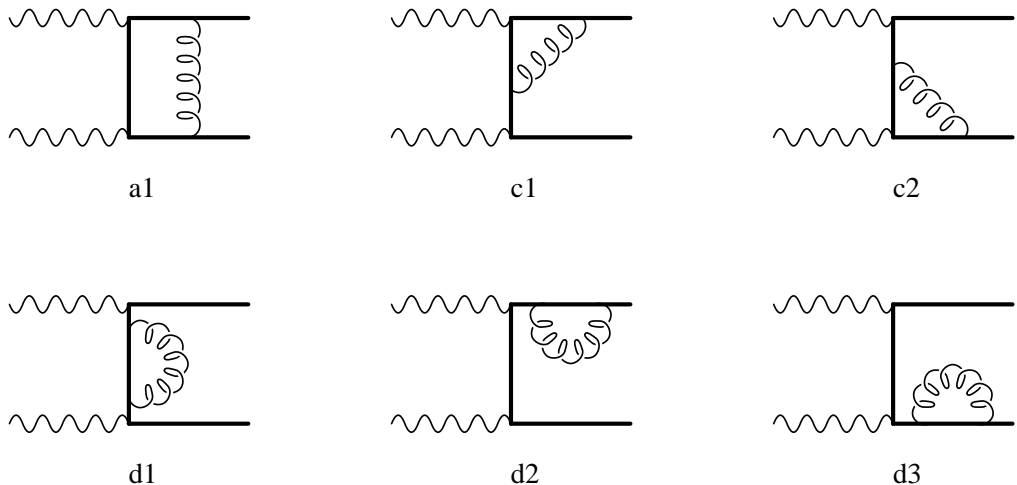}
\caption{\label{fig:gamgam}
One-loop Feynman diagrams contributing to the photon-fusion amplitude.
Wavy, curly, and solid lines represent photons, gluons, and heavy quarks,
respectively}
\end{figure*}

\section{\label{notation}
Notation
}

For consistency, we closely follow the notations of Ref.~\cite{KMC}.
The one-loop Feynman diagrams with $t$-channel topology relevant for
heavy-flavor production by the scattering of two on-shell photons
are depicted in Fig.~\ref{fig:gamgam}.
The $u$-channel diagrams are obtained from the depicted ones by crossing the
incoming photon lines.
Single-gluon radiation, which arises from the tree-level diagrams with a
gluon attached in all possible ways to the heavy-quark line,
contributes at the same order. 
We assign the four-momenta and helicities as
\begin{equation}
\label{gamgam}
\gamma(p_1,\lambda_1) + \gamma(p_2,\lambda_2)
                  \rightarrow Q(p_3) + \overline Q(p_4) + g(p_5) ,
\end{equation}
so that $p_1+p_2=p_3+p_4+p_5$, and have
$p_1^2=p_2^2=p_5^2=0$ and $p_3^2=p_4^2=m^2$, where $m$ is the quark mass.
We introduce the following Mandelstam variables:
\ba
s&=& (p_1+p_2)^2,
\nn\\
t&=& T-m^2=(p_1-p_3)^2-m^2,
\\
u&=& U-m^2=(p_2-p_3)^2-m^2,
\nn\\
s_2&=& S_2 - m^2 = (p_1 + p_2 - p_3)^2 - m^2 = s + t + u ,\nn
\qquad
\ea
so that $s_2 = 0$ in the soft-gluon limit.
Introducing 
\be
v= 1 + \frac{t}{s}, \qquad   w = \frac{-u}{s+t}
\ee
we may write
\be
t = -s (1-v),  \qquad  u = -svw,  \qquad  s_2 = sv(1-w).
\ee

In Ref.~\cite{KMC}, the four-momentum of the gluon was integrated out, 
the squared amplitudes were summed over the spins and colors of the 
final-state heavy quarks and averaged over the spins of the initial photons.
The differential cross sections $d\Delta\sigma/dvdw$ and $d\sigma/dvdw$ for
the polarized  and unpolarized cases were presented analytically, while the
total cross sections $\Delta\sigma$ and $\sigma$ were calculated numerically.
In Ref.~\cite{KM}, these differential cross sections were further integrated
to obtain fully analytical result for all the terms proportional to
$\delta(1-w)$, i.e., those that multiply the leading-order term.
However, the hard-bremsstrahlung contributions were left out, for which 
a suitable set of parametrizations were constructed. 
It is the aim of the present work to analytically integrate these
remaining contributions, e.g.\ the expression
in Eq.~(30) of Ref.~\cite{KMC}, except for its last term, 
proportional to $d\sigma_{\rm LO}/dvdw$.

\section{\label{evaluation}
Evaluation
}

In order to obtain the analytical result for the 
total cross section, one has to perform double integrations
over the variables $v$ and $w$, as was already mentioned in 
Sec.~\ref{notation}.
The explicit forms of the relevant integrals $I_i$ ($i=1,\dots,16$) are given
in Appendix~C of Ref.~\cite{KMC}.
However, their direct analytical evaluation turns out to be very 
complicated in general and even an unfeasible task in some cases. 
The integrals $I_i$ contain logarithms with square roots in their arguments, 
and their coefficient functions also depend on the integration variables.
In several cases where direct integration is possible, one obtains expressions 
in terms of the generalized Nielsen polylogarithms \cite{Devoto:1983tc}. 
These polylogarithms, however, contain long and complicated arguments 
that look unnatural, so that we decided to find some other
universal representation that would be valid for all the integrals
under consideration.

In fact, we made use of another approach to obtain the results.
The essence of our method consists in obtaining the integrated 
result from its expansion over the variable it depends on, 
as well as in the knowledge of the basis functions
entering the integrated result. In the past, such an approach
was used in Ref.~\cite{Fleischer:1998nb} for vertex- and 
propagator-type two-loop diagrams and was also applied to some other 
problems (see, e.g., Ref.~\cite{Kalmykov:2000qe}).

In our case, the result depends on the single variable $m^2/s$.
We find it, however, more convenient to set up the expansion
in the heavy-quark velocity
\be
\beta = \sqrt{1-\frac{4m^2}{s}} .
\label{velocity}
\ee
The procedure for obtaining the required expansions of the 
double integrals in the variable $\beta$, by first expanding and then
integrating 
Eq.~(30) of Ref.~\cite{KMC}, was already
discussed in detail in Ref.~\cite{KM} and will not be addressed here. 
We only mention that, in Ref.~\cite{KM}, only the first 11 terms of the 
expansions 
were obtained, which was all one could achieve at that time with available 
computer hardware resources.
For our present purposes, we needed to greatly enlarge the depths of the 
expansions.
Although this appears to be a straightforward task at first sight, it turned
out to be a major technical challenge in practice.
We actually needed hundreds of expansion terms to be able to rebuild the final 
integrated results.
For a given integral, the number of expansion terms is, of course, directly
connected to the number of functions that makes up our basis.
Thus, the main
problems were, on the one hand, to define the smallest possible basis and,
on the other hand, to
obtain sufficiently many terms of the expansion. 
Analyzing already integrated parts of the cross section presented in
Ref.~\cite{KM} and taking into account observations made in a number of
previous phenomenological studies, we chose our set of basis functions to be
the complete set
of harmonic polylogarithms of Remiddi and Vermaseren \cite{Remiddi:1999ew}.
Further detailed investigation revealed, however, that 
harmonic polylogarithms alone
are not sufficient, and that some nonharmonic functions should be added to
the basis, as will be explained below. These functions fall into the class
of multiple polylogarithms \cite{multilogarithm}.

It is well known that Feynman amplitudes satisfy linear 
differential equations (see, e.g., Ref.~\cite{Kotikov}). 
In order to establish the structure of the results, we found it very
convenient to consider homogeneous differential equations for the various
integrals $I_i$, which are of the form
\be
  \sum\limits_{n=0}^k P_n^{(i)}(\beta) \frac{d^n}{d\beta^n} I_i(\beta) = 0  \,,
\label{difur}
\ee
where $P_n^{(i)}(\beta)$ are some polynomials and $k$ is the order of the 
homogeneous differential equation.
Having typically 150--200 coefficients of an expansion in $\beta$, we were
able to 
establish the differential equations of the above type for each of the $I_i$
functions. 
As a result, we found 
that the degrees of the polynomials $P_n^{(i)}(\beta)$ never exceed 14 
and that the orders of the differential equations do not exceed 7. 
After having obtained the polynomials, 
one can try to solve the homogeneous differential equations by 
using the linear ansatz
\be
I_i = \sum\limits_{l} a_l^{(i)} F_l, 
\ee 
where the sum runs over all the basis functions. 
We remark that the first coefficients of the original $\beta$ expansions 
serve as boundary conditions for our differential equations.
Substitution 
of such an ansatz into Eq.~(\ref{difur}) leads to an algebraic
system of linear equations.

As already mentioned, not all the integrals can be given in terms
of harmonic polylogarithms. In particular, this was the case for
the integral $I_8$ of Ref.~\cite{KMC}, 
which is one of the most complicated ones.
We explicitly integrated the doubly differential distribution 
associated with this function. The details are presented in Appendix~A. 
Nevertheless, such a direct integration would be rather tedious for a majority 
of our functions, and the changes of variables described in Appendix~A are not
universal and, therefore, not applicable to the other integrals.
Another integral that cannot be expressed in terms of harmonic
polylogarithms is $I_6$.

Originally, our ansatz contained more 
than 100 basis functions. 
To work with such an ansatz, we needed about 1000 expansion coefficients
in the $\beta$ series. 
Finally, after some analysis, we constructed a final set of 21 basis
functions.
They are harmonic polylogarithms, except for three, which are discussed in
Sec.~\ref{results}.
With this set of basis functions,
the number of linear equations required 
varies from several tens to a couple of hundreds, depending on the
function $I_i$ considered.
Typically, one needs about 150--200 coefficients of the 
$\beta$ expansion to find the solution for the double integral $I_i$.

\section{\label{results}
Integrated results
}

The unpolarized and polarized cross sections are defined in terms of
$\sigma_{\lambda_1 \lambda_2}$ as
\ba
\sigma_{\rm unp}& =& \frac{1}{2} \left( \sigma_{++} + \sigma_{+-}\right) ,
\nn\\
\sigma_{\rm pol}& =& \frac{1}{2} \left( \sigma_{++} - \sigma_{+-}\right) .
\ea
We parametrize the total cross section in terms of the polarization
of the initial beams as
\begin{eqnarray}
\label{sigmadef}
  \sigma = 
      \frac{2-j}{2} \sigma_{++} + \frac{j}{2} \sigma_{+-} \,,
\end{eqnarray}
where $j=1-\langle \lambda_1\lambda_2 \rangle$ involves the average product
of the photon helicities $\lambda_1$ and $\lambda_2$. 
According to Eq.~(\ref{sigmadef}),
$j=1$ corresponds to the unpolarized cross section $\sigma_{\rm unp}$, while
$j=0$ and $j=2$ correspond to $\sigma_{++}$ and $\sigma_{+-}$, respectively.

At NLO, Eq.~(\ref{sigmadef}) can be written as
\begin{eqnarray}
  \sigma = \frac{2\pi\alpha^2 e_Q^4 N_c}{s}
    \left[ f^{(0)}(j,\beta) + C_F \frac{\alpha_s}{\pi} f^{(1)}(j,\beta)
 \right],
\label{eq:final}
\end{eqnarray}
where $e_Q$ is the fractional electric charge of the heavy quark $Q$, 
$N_c$ the number of colors,
and $\alpha$ the fine-structure constant.

The Born result is well known and reads
\begin{eqnarray}
\label{Born}
  f^{(0)}(j,\beta)& =&  2\beta + 2\beta^3 - 6j\beta
    - ( 1 + 2j - \beta^4 ) \ln\frac{1-\beta}{1+\beta} .
\nonumber\\
&&
\end{eqnarray}

The NLO result can be presented as a linear combination of universal basis
functions $F_i$,
\be
f^{(1)}(j,\beta) = \sum^{21}_{i=1} \, c^{(j)}_i(\beta)  F_i(\beta) ,
\label{res}
\ee
where all the $j$ dependence resides in the coefficients $c^{(j)}_i$.
The coefficients $c^{(j)}_i$ are given in Appendix~\ref{coeffs}.
The choice of the basis functions $F_i$ is not unique.
We choose the 21 basis functions $F_i$ as follows:
\ba
F_1 &=& 1, \qquad F_2 = H_1, \qquad F_3 = H_{-1},  \qquad F_4 = H_{1,1},\nonumber \\
F_5 &=& H_{-1,1}, \qquad  F_6 = H_{1,-1}, \qquad  F_7 = H_{-1,-1},\nonumber \\
F_8 &=& \Li_2(\beta) - \Li_2(-\beta), \qquad  F_9 = H_{1,1,1}, \nonumber \\
F_{10} &=& H_{-1,1,1}, \qquad  F_{11} = H_{1,-1,1}, \qquad  F_{12} = H_{-1,-1,1}, 
\nonumber \\
F_{13} &=& H_{1,1,-1}, \qquad  F_{14} = H_{-1,1,-1}, \quad  F_{15} = H_{1,-1,-1}, 
\nonumber \\
F_{16} &=& H_{-1,-1,-1},                 \nonumber   \\
F_{17} &=& 2 \int^{\beta}_0 \frac{db}{1-b^2}
\, \left[ \Li_2(b) - \Li_2(-b)\right],\nonumber \\
F_{18} &=& \int^{\beta}_0 \frac{db}{b} \, \ln^2\frac{1-b}{1+b},
\nonumber \\
F_{19} &=& 2 \int^{\beta}_0 \frac{bdb}{5-b^2} \, 
\left[\frac{1}{2} \ln^2(1+b)
- \frac{1}{2} \ln^2(1-b)\right. \nonumber \\
&&{}+\left. \Li_2\left(\frac{1+b}{2}\right) - 
\Li_2\left(\frac{1-b}{2}\right) + \ln 2 \ln \frac{1-b}{1+b} 
\right],\nonumber \\
F_{20} &=& 2 \int^{\beta}_0 \frac{bdb}{3+b^2} \,  \left(
\ln^2 \frac{1+b}{2} - \ln^2\frac{1-b}{2} \right),
\nonumber \\
F_{21} &=& \beta \, \left[A_1(\beta) - A_2(\beta) \right].
\label{res1}
\ea
The $H$ functions appearing in Eq.~(\ref{res1}) are the so-called 
harmonic polylogarithms, defined as 
\ba
H_{\pm 1}(\beta) &=&  \int^{\beta}_0 \frac{db}{1\mp b} = \mp \ln (1\mp \beta),
\nonumber \\
H_{\pm 1, a, ...}(\beta) &=&  \int^{\beta}_0 \frac{db}{1\mp b} H_{a, ...}(b);
\label{res2}
\ea
$\Li_2$ is the dilogarithm, defined below Eq.~(\ref{eq:s}); and
the functions $A_1$ and $A_2$ have the following compact one-fold integral
representations:
\ba
A_1(\beta) &=& \int^{\beta}_0 \, db \, \frac{\ln(1-b^2)}{2-\beta^2 -b^2}
\ln \frac{1-b^2}{1-\beta^2},
\nonumber \\
A_2(\beta) &=& \int^{\beta}_{-\beta} \, db \, \frac{\ln(1+b)}{2-\beta^2 -b^2}
\left[2\ln (1+b)\right. \nonumber \\
&&{}-\left.\ln (1+\beta^2+2b)\right].
\label{res3}
\ea

We observe that only three basis functions
$F_{19}$, $F_{20}$, and $F_{21}$ of Eq.~(\ref{res1}) are not expressible in
terms of harmonic polylogarithms~(\ref{res2}).
We remark that
the basis function $F_{20}$ arises only from the virtual 
part of the cross section.

We note that all the basis functions $F_i$ of Eq.~(\ref{res1})
can also be expressed via generalized Nielsen polylogarithms,
\begin{equation}
\Si_{n,p}(y)=\frac{(-1)^{n+p-1}}{(n-1)!\,p!}
\int\limits_0^1\frac{dt}{t} \, \ln^p(1-ty) \, \ln^{n-1}t
\label{eq:s}
\end{equation}
with $n+p=2,3$ and complicated arguments. 
Special cases include the polylogarithm of order $n$,
$\Li_n(y)=\Si_{n-1,1}(y)$, and Riemann's zeta function 
$\zeta_n=\zeta(n)=\Li_n(1)$ \cite{Lewin,Devoto:1983tc}.
We rewrite the functions $F_i$ in terms of the standard 
generalized Nielsen polylogarithms in Appendix~\ref{functions}.
We note in passing that all the $F_i$ functions can be
expressed in terms of multiple polylogarithms $L_{1,1,1}$ of 
depth and weight 3 \cite{multilogarithm} with simple linear arguments.

To verify our analytical results, we compared the numerical values
for the function $f^{(1)}$ of Eq.~(\ref{eq:final}) produced by our
{\sc Mathematica} program 
in the polarized and unpolarized cases with Table~1 of Ref.~\cite{KM}.
There, the values for $f^{(1)}$
are presented as functions of the variable
\be
z=\frac{\sqrt{s}}{2m} = \frac{1}{\sqrt{1-\beta^2}}.
\ee
We found agreement on the level of better
than one part in $10\,000$. 
Next, we compared our numbers with those for $z=1,2,3,4,5,10$
presented in Table~1 of Ref.~\cite{Mirkes} dealing 
with the unpolarized case.
The agreement was at the order of 
one part in $10\,000$ or better. 
Finally, we also compared our numbers with the corresponding values
for $f^{(1)}(++)$ and $f^{(1)}(+-)$ from Ref.~\cite{Jikia}. 
Generally, we were in good agreement; however, we found deviations for
$f^{(1)}(++)$ by about 3\% at some values of $z$.

The present results form an Abelian subset
of the non-Abelian gluon-induced NLO contributions to heavy-quark
pair production.
Recently, the total cross section of this subprocess was calculated
analytically for unpolarized gluons in Ref.~\cite{Czakon:2008ii} using a
completely different approach. 
By modifying the color structures, it is possible to extract the 
unpolarized $\gamma\gamma$ cross section from their result.
Comparing both numerically and analytically (after expanding in $\beta$), we
find complete agreement. 
Specifically, three nonharmonic functions $F_1(x)$, $F_2(x)$, and $F_3(x)$
appearing in Eqs.~(13)--(15) of Ref.~\cite{Czakon:2008ii} can be expressed 
as linear combinations of our functions $F_{19}$, $F_{20}$, and $F_{21}$.
For instance, for the most complicated function $F_3(x)$, one has
\be
F_{21}(\beta) = \frac{8}{15} \frac{1-x}{\sqrt{1+6x+x^2}} F_3(x),
\ee
where $x=(1-\beta)/(1+\beta)$.

\section{\label{summary}
Conclusions
}

We presented, in analytic form, the integrated total cross sections of
heavy-quark production in polarized and unpolarized $\gamma \gamma$ collisions
at NLO in QCD.
The result is written as a sum over bilinear products of $j$-dependent
coefficient functions and $j$-independent basis functions, where $j$ denotes
the total angular momentum of the photons.

We checked our analytical results by reproducing, with great accuracy, 
all the numerical values listed in the relevant tables of
Refs.~\cite{KM,Mirkes}.
Furthermore, we established agreement with the analytic NLO result for the
total cross section of heavy-quark production via $gg$ fusion, obtained just
recently in Ref.~\cite{Czakon:2008ii}, by taking the Abelian limit.

Using the backscattering technique, it is straightforward to obtain 
polarized-photon beams of high intensity at the $\gamma \gamma$
option of the ILC by colliding low-energy laser light with
polarized electron and positron beams.

Of some concern are resolved-photon contributions.
On the one hand, the unpolarized cross sections of the contributing
subprocesses were already presented 
in Ref.~\cite{Czakon:2008ii} and the polarized ones may be deduced, e.g.,
from Ref.~\cite{Bojak}.
On the other hand, such contributions can be suppressed by operating
close to the production threshold.
In fact, we infer from Ref.~\cite{Zerwas} that, in the case of $b$-quark
production close to threshold, the resolved contribution only makes up
a fraction of a percent of the full cross section.
Resolved contributions may also be reduced by identifying outgoing jets
collinear to one of the photon beams, which are a signature of resolved-photon
events.
One can also require that the energy deposited in the detectors be equal 
to the total beam energy in order to account for missed jets of the type
mentioned above. 
From the experimental side, we are assuming only that heavy-quark events can
be clearly identified.

Our computer program evaluates the total cross
sections presented here in less than a second.
It is publicly available \cite{offering}
and uses the program package HPL \cite{Maitre}. 
Being implemented in {\sc Mathematica}, it does not allow for calculations
with arbitrary precision.
However, with some additional technical modifications, arbitrary 
precision could be achieved.

\begin{acknowledgments}
We thank A.I.~Davydychev, M.Yu.~Kalmykov, and O.V.~Tarasov for useful
discussions.
We also thank M.~Czakon and A.~Mitov for assistance in comparing their
results \cite{Czakon:2008ii} with ours.
The work of B.A.K. was supported in part by the German Federal Ministry for
Education and Research BMBF through Grant No.~05~HT6GUA.
The work of A.V.K. was supported in part by the German Research Foundation DFG
through Mercator Guest Professorship No.\ INST~152/465--1,
by the Heisenberg-Landau Programme through Grant No.~5,
and by the Russian Foundation for Basic Research through Grant
No.~08--02--00896--a.
The work of Z.V.M. was supported in part by the DFG through Grant
No.~KN~365/7--1 and by the Georgia National Science Foundation through Grant
No.~GNSF/ST07/4--196.
The work of O.L.V. was supported by the Helmholtz Association through Grant
No.~HA--101.
\end{acknowledgments}


\appendix

\section{Integral transformations}
\label{I8integral}

The contribution proportional to the integral $I_8$ of Ref.~\cite{KMC} can be 
represented in the form
\ba
\frac{I_8}{N} = \int^{v_2}_{v_1} dv  \int^1_{w_1} 
\frac{v dw}{\sqrt{x_8}} \, \ln
\frac{\sqrt{x_8}+v\overline{w}}{\sqrt{x_8}-v\overline{w}}\,  e_8(v,w),
\label{I8.1}
\ea
where $N$ is a known normalization constant and
\ba
v_1&=&\frac{1-\beta}{2},\quad
v_2=\frac{1+\beta}{2},\quad 
w_1 = \frac{a}{v(1-v)}, 
\nonumber \\ 
\overline{w}&=&1-w,\quad
x_8=v^2\overline{w}^2 + 4av(1-vw),\quad
a=v_1v_2.
\nonumber\\
&& \label{I8.2}
\ea
In the polarized case, the coefficient function $e_8$ is simply replaced by
$\Delta e_8$. The actual expressions for $\Delta e_8$ and $e_8$
may be found in Eqs.~(B6) and (B8) of Ref.~\cite{KMC}, respectively.

It is convenient to change the order of integrations as
\ba
\int^{v_2}_{v_1} dv  \int^1_{w_1} dw  \longrightarrow   
\int^1_{1-\beta^2}  dw \int^{\frac{1+t}{2}}_{\frac{1-t}{2}}  dv  ,
\label{I8.3}
\ea
where $t^2= 1- 4a / w$.
Furthermore, $x_8$ defined in Eq.~(\ref{I8.2}) can be represented as
\ba x_8 = v^2 y_8,\qquad
y_8 = \overline{w}^2 + 4a\left(\frac{1}{v}-w\right),
\label{I8.5}
\ea
so that we may substitute
\ba
\frac{v}{\sqrt{x_8}}  \ln
\frac{\sqrt{x_8}+v\overline{w}}{\sqrt{x_8}-v\overline{w}} =
\frac{1}{\sqrt{y_8}}  \ln 
\frac{\sqrt{y_8}+\overline{w}}{\sqrt{y_8}-\overline{w}} .
\label{I8.6}
\ea
in Eq.~(\ref{I8.1}). 

Clearly, the ``natural'' replacement 
$\xi=1/v$ renders $y_8$ just linearly dependent on the new variable $\xi$.
As a consequence, the integral in Eq.~(\ref{I8.1}) will be transformed as
\ba
\frac{I_8}{N} &=& \int^1_{1-\beta^2} dw\, R(w), 
\nonumber \\
R(w) &=& \int^{\frac{w}{2a}(1+t)}_{\frac{w}{2a}(1-t)}
\frac{d\xi}{\sqrt{y_8}}  \ln
\frac{\sqrt{y_8}+\overline{w}}{\sqrt{y_8}-\overline{w}} \, 
\overline{e}_8(x,w) ,
\label{I8.7}
\ea
where
$\overline{e}_8 \equiv e_8/\xi^2$.

The next step is to replace the integration variable $\xi$ by 
the new integration variable $z=\sqrt{y_8}$, so that the square root is removed
from the logarithm. 
Thus, one obtains
\ba
R(w) = \frac{1}{2a} \int^{\sqrt{y_+}}_{\sqrt{y_-}} 
dz \, \ln 
\frac{z+\overline{w}}{z-\overline{w}} \, \overline{e}_8(z,w),
\label{I8.9}
\ea
where
$y_{\pm} = 2w(1\pm t-2a)+\overline{w}^2$.

It is then convenient to split $R(w)$ into the two parts as
\ba
R(w) &=& R_+(w)-R_-(w), \label{I8.11} \\
R_\pm(w) &=& \frac{1}{2a} \int^{\sqrt{y_+}}_{\sqrt{y_-}}
dz \, \ln (z\pm\overline{w}) \, \overline{e}_8(z,w) \nonumber \\
&=& \frac{1}{2a} \int^{\sqrt{y_+}\pm\overline{w}}_{\sqrt{y_-}\pm\overline{w}}
dz_\pm \, \ln z_\pm \, \overline{e}_8(z_\pm,w), \label{I8.12}
\ea
where $z_\pm=z\pm\overline{w}$,
which induces a corresponding split of the original integral $I_8/N$:
\begin{equation}
\frac{I_8}{N} = I_8^{(+)}-I_8^{(-)} .
\label{I8.14}
\end{equation}
After exchanging the order 
of integrations and performing some algebraic manipulations, we obtain
\ba
I_8^{(\pm)} &=& \frac{1}{2a} 
 \int^{1+\beta}_{1-\beta} 
dz_\pm \, \ln z_\pm \int^1_{w_\pm} 
dw \, \overline{e}_8(z_\pm,w), \nonumber \\
&=&  \frac{1}{2a} 
 \int^{\beta}_{-\beta}
dr_\pm \, \ln (1\pm r_\pm) \int_0^{\overline{w}_\pm} 
d\overline{w} \, \overline{e}_8(r_\pm,\overline{w}),\qquad
\label{I8.15}
\ea
where
\ba
z_\pm&=&1\pm r_\pm, \qquad w_\pm = \frac{(1-r_\pm)^2}{2(1-r_\pm-2a)},
\nonumber\\ 
\overline{w}_\pm &=& \frac{\beta^2-r_\pm^2}{2(1-r_\pm-2a)} .
\label{I8.16}
\ea

It turns out the function $\overline{e}_8$, when expressed in terms of the new
variables $r_\pm$, is greatly simplified, and so is the integration
over the variable $w$.
Performing the integrals in Eq.~(\ref{I8.15}),
most of the terms contained in $\overline{e}_8$ yield harmonic polylogarithms
$H$ and generalized Nielsen polylogarithms $\Si_{a,b}$. 
Only the most complicated terms of $\overline{e}_8$ lead to
the structures $A_1(\beta)$ and $A_2(\beta)$ in Eq.~(\ref{res3}).

\section{Coefficients}
\label{coeffs}

Here, we list the coefficients $c_i^{(j)}$ appearing in Eq.~(\ref{res}).
They read
\ba
c^{(j)}_1 &=& -19 \beta + 41 \beta^3 + \left(1+\frac{3}{2} \beta + 2\beta^2 -
\frac{1}{2} \beta^3 + \beta^4\right) \pi^2
\nonumber \\
&&{}- 4\beta (1+2\beta^2) \ln (2\beta^2)
\nonumber \\
&&{}+ j  \biggl[-32 \beta -\frac{3}{4} \left(4 +\beta +4\beta^2 - \frac{3}{4} 
\beta^3\right)
\pi^2
\nonumber \\
&&{}+ 4\beta \left(7 - \frac{2}{3 + \beta^2}\right) \ln 2 + 24\beta \ln \beta\biggr],
\nonumber \\
c^{(j)}_{2,3} &=& -\frac{9}{2} \mp2 \beta -13 \beta^2 \mp4 \beta^3 -\frac{25}{2} \beta^4
+ \frac{80}{5 - \beta^2}
\nonumber \\
&&{}+ \frac{1}{4\beta} (2- \beta + 2\beta^2 +2 \beta^3 -2 \beta^4-\beta^5-2\beta^6) \pi^2
\nonumber \\
&&{}+ 2 \left(7+\beta^2+4\beta^4-\frac{24}{3 + \beta^2}\right) \ln 2 + 8\beta^2 (1+\beta^2) 
\ln\beta
\nonumber \\
&&{}+ j  \biggl\{9\pm14 \beta +7\beta^2 - \frac{40}{5 - \beta^2} 
+ \frac{4(3\mp\beta)}{3 + 
\beta^2}
\nonumber \\
&&{}+ \frac{1}{8\beta} (8 +\beta +8\beta^2 -2 \beta^3+\beta^5) \pi^2
\nonumber \\
&&{}- 2 \left[21 + 5\beta^2 -\frac{48}{3 + \beta^2}+\frac{48}{(3 + \beta^2)^2}\right] \ln 2
\nonumber \\
&&{}- 4 (5+3\beta^2)\ln \beta\biggr\},
\nonumber \\
c^{(j)}_{4,7} &=& \mp18 -9\beta \pm10 \beta^2 +3 \beta^3 \mp6 \beta^4
\pm \frac{240}{5 - \beta^2} \mp \frac{320}{(5 - \beta^2)^2}
\nonumber \\
&&{}\mp \frac{48}{3 + \beta^2} +
\frac{1}{2\beta} (8\pm3\beta+8\beta^2\mp6\beta^3-8\beta^4\pm3\beta^5
\nonumber \\
&&{} -8\beta^6) \ln 2 +
\frac{4}{\beta}(1+\beta^2-\beta^4-\beta^6) \ln \beta
\nonumber \\
&&{}+ j \biggl[\mp\frac{39}{2}+\frac{9}{2}\beta \pm\frac{39}{2}\beta^2
-\frac{3}{2}\beta^3
\nonumber \\
&&{}\pm
\frac{160}{(5 - \beta^2)^2} \mp \frac{160}{5 - \beta^2} \mp
\frac{96}{(3 + \beta^2)^2} \pm \frac{96}{3 + \beta^2}
\nonumber \\
&&{}+ \frac{1}{4\beta} (32 \pm29\beta + 8\beta^2 \pm 26\beta^3  + 8\beta^4 \mp 15\beta^5
) \ln 2
\nonumber \\
&&{}+ \frac{8}{\beta} (1+\beta^2)\ln \beta\biggr],
\nonumber \\
c^{(j)}_{5,6} &=& \
\pm32 -9\beta \mp8 \beta^2 +3 \beta^3 \pm14 \beta^4
\mp \frac{240}{5 - \beta^2} \pm\frac{320}{(5 - \beta^2)^2}
\nonumber \\
&&{}+ \frac{1}{2\beta} (8\mp3\beta+8\beta^2\pm6\beta^3-8\beta^4\mp3\beta^5-8\beta^6) \ln 2
\nonumber \\
&&{}+
\frac{4}{\beta}(1+\beta^2-\beta^4-\beta^6) \ln \beta
\nonumber \\
&&{}+ j \biggl[\mp\frac{45}{2}+\frac{9}{2}\beta \mp\frac{59}{2}\beta^2
-\frac{3}{2}\beta^3 \mp
\frac{160}{(5 - \beta^2)^2} \pm \frac{160}{5 - \beta^2}
\nonumber \\
&&{}+ \frac{1}{4\beta} (32 \mp29\beta + 8\beta^2 \mp 26\beta^3  + 8\beta^4 \pm 15\beta^5
) \ln 2
\nonumber \\
&&{}+ \frac{8}{\beta} (1+\beta^2)\ln \beta\biggr],
\nonumber \\
c^{(j)}_8 &=& -4 -16 \beta^2 -12 \beta^4 + 8j (
4 +3\beta^2 ),
\nonumber \\
c^{(j)}_{9,16} &=& 21 \mp \frac{1}{\beta} \mp\beta
+18\beta^2 \pm\beta^3-7 \beta^4 \pm \beta^5
\nonumber \\
&&{}+  \frac{j}{4\beta}  (\mp8 -\beta \mp 26\beta^2 - 34\beta^3 \pm 6\beta^4 + 7\beta^5
),
\nonumber \\
c^{(j)}_{10,15} &=& -18 \mp \frac{1}{\beta} \mp\beta
-24\beta^2 \pm\beta^3+10 \beta^4 \pm \beta^5
\nonumber \\
&&{}+  \frac{j}{4\beta}   (\mp8 -5\beta \mp 26\beta^2 + 46\beta^3 \pm 6\beta^4
 - 13\beta^5
),
\nonumber \\
c^{(j)}_{11,14} &=& - \frac{33}{2} \pm \frac{5}{\beta} \pm5\beta
-27\beta^2 \mp5\beta^3+\frac{23}{2} \beta^4 \mp5 \beta^5
\nonumber \\
&&{}+  \frac{j}{2\beta}   (\pm20 +12\beta \pm 17\beta^2 + 36\beta^3 \pm\beta^4
 - 14\beta^5
),
\nonumber \\
c^{(j)}_{12,13} &=&  \frac{39}{2} \pm \frac{5}{\beta} \pm5\beta
+21\beta^2 \mp5\beta^3-\frac{17}{2} \beta^4 \mp5 \beta^5
\nonumber \\
&&{}+  \frac{j}{2\beta}  (\pm20 -15\beta \pm 17\beta^2 - 30\beta^3 \pm\beta^4
+ 11\beta^5
),
\nonumber \\
c^{(j)}_{17} &=& -\frac{6}{\beta} -6\beta +6 \beta^3 + 6\beta^5 - \frac{12j}{\beta}
(1 +\beta^2),
\nonumber \\
c^{(j)}_{18} &=& \frac{c^{(j)}_{17}}{3},
\nonumber \\
c^{(j)}_{19} &=& 20 + 30\beta^2 - 10\beta^4 - \frac{5j}{4}(
7 + 14\beta^2 - 5\beta^4),
\nonumber \\
c^{(j)}_{20} &=& -\frac{3}{4} +\frac{3}{2} \beta^2 -\frac{3}{4} \beta^4 -
\frac{3j}{8}(
15 +6 \beta^2 -5\beta^4),
\nonumber \\
c^{(j)}_{21} &=& - \frac{11}{\beta} -15\beta
+11\beta^3- \beta^5 +  \frac{2j}{\beta}
(5 -\beta^2).
\label{resco}
\ea

\section{Basis functions}
\label{functions}

As was already mentioned in Sec.~\ref{results}, all the $F_i$ functions 
in Eq.~(\ref{res1}) can be written in terms of $\Si_{n,p}$ functions with
$n+p=2,3$ and some complicated arguments. 
The functions $F_i$ with $i=1,\ldots,16$ are written in terms of the standard 
harmonic polylogarithms of Remiddi and Vermaseren \cite{Remiddi:1999ew}, and
their representations in terms of Nielsen polylogarithms may be
found in Ref.~\cite{Remiddi:1999ew}.
The functions $F_i$ $(i=17,\ldots,21)$ have the following forms:
\ba
F_{17} &=& -\ln x \left[ \Li_2(\beta) - \Li_2(-\beta) \right]- F_{18},
\nonumber \\
F_{18} &=& \ln \beta \ln^2 x - 3\zeta_3 + 2 \left[ \vp_1(1,x)- \vp_1(-1,x)
 \right],
\nonumber \\
F_{19} &=& F_{19}^{(1)} +  2 F_{19}^{(2)}+ 2 F_{19}^{(3)},
\nonumber \\
F_{19}^{(1)} &=& -\ln(5-b^2) 
\left[2\Li_2(-x)+\zeta_2 + 2\ln x \ln (1+x)
\vphantom{\frac{1}{2}}\right.
\nonumber \\
&&{}-\left.
\frac{1}{2} \ln^2 x \right],
\nonumber \\
F_{19}^{(2)} &=& 
\ln 2\left(
 \ln^2\frac{1+\beta}{2}-\ln^2\frac{1-\beta}{2}
\right) 
 \nonumber \\
&&{}+  \vp_1\left(z_1^-,\frac{1+\beta}{2}\right)- 
\vp_1\left(z_1^-,\frac{1-\beta}{2}\right) 
 \nonumber \\
&&{}+ 
\vp_1\left(-z_1^+,\frac{1+\beta}{2}\right)- 
\vp_1\left(-z_1^+,\frac{1-\beta}{2}\right) ,
 \nonumber \\
F_{19}^{(3)} &=& 
\ln 4\biggl[2\Li_2(-x)+\zeta_2 + \frac{1}{2} \ln^2 x \biggr]
 \nonumber \\
&&{}- \frac{5}{2} \zeta_3 + 8 \Si_{1,2}(-x) - 2 \vp_1(-1,x)
 \nonumber \\
&&{}+ 2 \vp_2(-z_2^-,x)- 2 \vp_2(-z_2^+,x),
 \nonumber \\
F_{20} &=& F_{20}^{(1)} +  2 F_{20}^{(2)},
\nonumber \\
F_{20}^{(1)} &=& \ln\frac{3+\beta^2}{4}
\left(
 \ln^2\frac{1+\beta}{2}-\ln^2\frac{1-\beta}{2}
\right),\nonumber \\
F_{20}^{(2)} &=& 
\frac{1}{9} \vp_1\left(-1,{\left(\frac{1+\beta}{2}\right)}^3\right)
- \frac{1}{9} \vp_1\left(-1,{\left(\frac{1-\beta}{2}\right)}^3\right)
\nonumber \\
&&{}- \vp_1\left(-1,\frac{1+\beta}{2}\right) + 
\vp_1\left(-1,\frac{1+\beta}{2}\right),
\nonumber \\
F_{21} &=& \frac{\beta}{2d}\left[
\ln \frac{1-\beta^2}{(d-1)^2}\, F_{21}^{(1)} 
+  F_{21}^{(2)} \right],
\nonumber \\
F_{21}^{(1)} &=&
\ln \delta \ln \rho - \frac{1}{2} \ln^2 \rho 
+\Li_2 (\delta \rho) -\Li_2 \left(\frac{\delta}{\rho}\right)
\nonumber \\
&&{}
-\zeta_2 -  2\Li_2 (- \rho),
\nonumber \\
F_{21}^{(2)} &=&
\zeta_3 -5\Si_{1,2}(-\rho) + 4\Si_{1,2} \left(-\frac{1}{\rho}\right)
+2 \Si_{1,2}(\delta )
\nonumber \\
&&{}-4 \Si_{1,2} \Bigl(\delta \rho\Bigr) + 
2 \Si_{1,2} \left(\frac{\delta}{\rho}\right) + 3 \pi^2 \ln \rho
\nonumber \\
&&{}
-5  \vp_2(1,\rho) + \vp_2\left(1,\frac{1}{\rho}\right)
\nonumber \\
&&{}+ \re \left\{\Si_{1,2} \left(\frac{1}{\delta}\right)
-\Si_{1,2} \left(\frac{\rho}{\delta}\right)
+ 2 \Phi \left(-\delta,-\frac{1}{\delta}\right)
\right.\nonumber \\
&&{}
- 3 \Phi \left(-\rho \delta,-\frac{\rho}{\delta}\right)
+  \Phi \left(-\frac{1}{\rho \delta},-\frac{\delta}{\rho}\right)
\nonumber \\
&&{}-\left.5  \vp_2\left(-\frac{1}{\delta},\rho\right) + 
\vp_2\left(-\frac{1}{\delta},\frac{1}{\rho}\right) \right\} ,
\label{res18-21}
\ea
where
\ba
z_{1}^{\pm} &=& \frac{\sqrt{5}\pm 1}{2},\  z_{2}^{\pm} = 
\frac{3 \pm \sqrt{5}}{2}, \nonumber \\
d &=& \sqrt{2-\beta^2},\  \delta = \frac{d-1}{d+1},\
\rho = \frac{d-\beta}{d+\beta}.
\label{argu}
\ea
The functions $\varphi_1$ and $\varphi_2$ are defined as
\ba
\vp_1(\alpha,x) &=&  \int^1_x \frac{dy}{y} \ln y \ln(1-\alpha y) 
\nonumber \\ 
&=&
\Li_3(\alpha)-\Li_3(\alpha x)+\ln x \Li_2(\alpha x), \nonumber \\
\vp_2(\alpha,x) &=& \int^1_x \frac{dy}{y} \ln (1 + y) 
\ln(1+\alpha y) 
\nonumber \\ 
&=& \Phi(1, \alpha)-\Phi (x, \alpha x), 
\label{fun1}
\ea
where (see Eq.~(3.15.4) of Ref.~\cite{Devoto:1983tc})
\ba
\Phi (A,B)&=&  \int^1_0 \frac{dy}{y} \ln (1+Ay) \ln(1+By) 
\nonumber \\
&=& \Si_{1,2}(-A)+ \Si_{1,2}(-B)-\frac{1}{2} \ln^2 \frac{A}{B}
\ln(1+B) \nonumber \\
&&{} + \ln\frac{A}{B} \left[\Li_2\left(\frac{A-B}{A}\right)
-\Li_2\left(\frac{A-B}{A(1+B)}\right) \right] \nonumber \\
&&{} -\Si_{1,2}\left(\frac{A-B}{A}\right)
+\Si_{1,2}\left(\frac{A-B}{A(1+B)}\right)
\nonumber \\
&&{}-\Si_{1,2}\left(\frac{B-A}{1+B}\right).
\ea


\end{document}